# Variational calculation of heavy-light meson properties [*]

Terrence Draper and Craig McNeile [a]

[a]Department of Physics and Astronomy, University of Kentucky, Lexington, KY 40506, USA

We present a new method for the study of heavy-light mesons in the static approximation of lattice QCD which is optimally effective in isolating ground and excited states. With "MOST" (Maximal Operator Smearing Technique), the heavy quark is smeared at all possible positions relative to the light quark, subject to the constraint of cubic symmetry. With correlation functions constructed using this set as a variational basis, eigenstates of the transfer matrix are projected out at very small time separations, where statistical errors are small. We illustrate the utility of the method with preliminary results for the meson decay constant $f_B^{\text{stat}}$, binding energies and wave functions of the lowest states. The method produces dynamically-improved interpolating fields which can be used for matrix element calculations.

Lattice QCD can make important contributions to phenomenology by the reliable calculation of decay constants, mixing amplitudes and form factors of $B$-mesons. It is natural to study the effective theory where the heavy $b$ quark is approximated by a static color source [1] giving a simpler system with only one dynamical (valence) quark.

To satisfactorily extract hadronic ground state properties from lattice correlation functions of operators, it is imperative to choose operators which couple much more strongly to the ground state than to excited states. With the first attempts at a lattice calculation of $f_B^{\text{stat}}$ in the static approximation, it was realized that local operators were very poor choices. A second generation of attempts used operators which were spatially smeared; this had been done successfully for light-light mesons. Smearing over cubes, over walls, or with Gaussian or exponential functions improved matters by improving statistics, but were still unsatisfactory [1]. The exponential degradation of the signal-to-noise ratio with increasing Euclidean time, being much more pronounced for static-light than for light-light mesons, requires that better, more physical, smearing functions be used.

The use of a spinless relativistic quark model Hamiltonian to construct a basis set of trial smearing functions for a variational calculation was an important improvement [2]. We propose an alternative: that the variational calculation be done in a manner free of any such ansatz, allowing the QCD dynamics to determine the optimal smearing functions.

We use interpolating fields which smear the static quark $Q$ *relative* to the light quark $q$

$$\Phi_{s_i}(\vec{x},t) = \sum_{\vec{r}} s_i(\vec{r})\overline{Q}(\vec{x}+\vec{r},t)\Gamma q(\vec{x},t) \qquad (1)$$

and choose as a basis the set of $N = (L/2 + 1)(L/2 + 2)(L/2 + 3)/6$ independent "S-wave" smearing functions

$$s_i(\vec{r}) = \frac{1}{48}\sum_{R \in O} \delta_{\vec{r},R\vec{r}_i} \qquad (2)$$

(This generalizes for other representations of the cubic symmetry group $O$). For $L^3 = 20^3$ spatial sites, $N = 286$ is a manageably small number, and so the $N \times N$ correlation matrix

$$C_{ij}(t) = \sum_{\vec{x}} \langle 0 \mid \Phi_{s_i}(\vec{x},t)\Phi_{s_j}^\dagger(\vec{0},0) \mid 0 \rangle \qquad (3)$$

contains *all* of the information available from two-quark interpolating fields. Fixing the origin of the light quark propagator, taking advantage that the static-quark propagator depends only on a single spatial variable, and utilizing fast Fourier transforms allows the inexpensive computation of the

[*]Presented by T. Draper. This work is supported in part by the U.S. Department of Energy under grant number DE-FG05-84ER40154 and by the National Science Foundation under grant number STI-9108764.

*entire* correlation matrix. This is a unique luxury afforded in the static approximation and provides the maximal (valence quark) basis to exploit variational methods.

It is illuminating to use matrix notation, and to write the correlation matrix as

$$C(t) = \mathcal{A} e^{-\mathcal{M}t} \mathcal{A}^\dagger \qquad (4)$$

with $\mathcal{A}_{in} \equiv \langle 0 \mid \Phi_i(\vec{0},0) \mid n \rangle$. In particular, $\mathcal{A}_{11}$ is the coupling of the local current to the ground state $n = 1$. The pseudoscalar decay constant is determined from

$$\tilde{Z}_L \equiv \mathcal{A}_{11} = f_B^{\text{stat}} \sqrt{m_B/2} (\tilde{Z}_A^{\text{stat}})^{-1} a^{3/2} \qquad (5)$$

$\tilde{Z}_A^{\text{stat}}$, defined by $A_\mu^{\text{cont}} = \tilde{Z}_A^{\text{stat}} A_\mu^{\text{latt}} a^{-3/2}$, is the renormalization factor [5]. (The ~ on $\tilde{Z}_L$ and $\tilde{Z}_A^{\text{stat}}$ indicates that all $\kappa$-dependent normalizations are absorbed into the renormalization factor.)

However, the $\mathcal{A}$'s are not orthogonal, and cannot be determined directly upon diagonalizing the matrix at time slice $t_0 \geq 0$

$$C(t_0) = V(t_0) \Lambda(t_0) V^\dagger(t_0) \qquad (6)$$

where $\Lambda(t_0)$ is the diagonal matrix of eigenvalues. These are approximately equal to the eigenvalues of the transfer matrix; corrections are damped exponentially only after large $t_0$ by which time the signal-to-noise ratio has been degraded.

Kronfeld [3] and Lüscher and Wolff [4] have outlined procedures by which eigenstates of the transfer matrix can be estimated at early times; the formalism, adapted to our variational basis of all relative spatial separations of the quarks, is very effective in isolating eigenstates. Following Kronfeld, there exists an orthogonal matrix $R(t_0)$ which relates the eigenvalues $\Lambda_{nn}(t_0)$ and eigenvectors $V_{in}(t_0)$ of the correlation matrix to those of the transfer matrix:

$$\left(\mathcal{A} e^{-\mathcal{M} t_0/2}\right) = \left(V(t_0) \Lambda(t_0)^{+1/2}\right) R(t_0) \qquad (7)$$

$R(t_0)$ can be determined from the correlation matrix at later times $t > t_0 \geq 0$. Diagonalizing

$$D(t,t_0) = \Lambda(t_0)^{-1/2} V(t_0)^\dagger C(t) V(t_0) \Lambda(t_0)^{-1/2}$$
$$= R(t_0) e^{-\mathcal{M}(t-t_0)} R(t_0)^\dagger \qquad (8)$$

determines the amplitudes

$$\mathcal{A} = V(t_0) \Lambda(t_0)^{+1/2} R(t_0) e^{+\mathcal{M} t_0/2} \qquad (9)$$

from which the decay constant(s) can be calculated directly using eqn. 5, and determines $\mathcal{M}_{nn}$ (which for static-light mesons are the bare binding energies $\mathcal{E}_n^0$).

As a important bonus, one obtains

$$\mathcal{Z} = V(t_0) \Lambda(t_0)^{-1/2} R(t_0) e^{-\mathcal{M} t_0/2} \qquad (10)$$

which projects out eigenstates of the transfer matrix: $\mathcal{Z}^\dagger \mathcal{A} = 1$. Thus $\Phi' = \mathcal{Z}^\dagger \Phi$ are optimally improved interpolating fields, obtained dynamically,

$$\mathcal{Z}^\dagger C(t) \mathcal{Z} = e^{-\mathcal{M}t} \qquad (11)$$

We have suppressed, for clarity, factors of $n_i$, the number of spatial points equivalent to a given point $\vec{r}_i$ under the cubic symmetry group; these remain upon condensing the full $L^3 \times L^3$ correlation matrix down to size $N \times N$.

All two-quark interpolating fields do not provide a complete basis (due to multiquark and hybrid intermediate states), so the exact expression in eqn. 8 is corrected by exponentially damped terms, which however are expected to be smaller than those correcting the extraction of eigenvalues of the transfer matrix directly from the eigenvalues of the correlation matrix [4]. Thus for moderate values of $t$ and $t_0$ one expects eqns. 9 and 10 to be good approximations.

The smaller eigenvalues of the correlation matrix are noisy and contribute little to the largest eigenvalues of the transfer matrix. We truncate the correlation matrix to its highest $n_\lambda$ eigenvalues before the construction in eqn. 8 [3]. This introduces further exponentially damped corrections which are expected to be small and which are dealt with as above.

For this report, rather than use eqn. 9 directly to determine $f_B^{\text{stat}}$, we instead use the dynamically-improved interpolating fields $\Phi' = \mathcal{Z}^\dagger \Phi$ (obtained from diagonalizing $D(3,2)$ with $n_\lambda = 5$), in the conventional way to determine $\tilde{Z}_L$ from fits to local-smeared and smeared-smeared correlations.

Quark propagators were computed for $\kappa = 0.152, 0.154$, and $0.155$ with the Wilson fermionic action in a background of 32 quenched (Coulomb gauge-fixed) configurations (selected every 1000 pseudo-heat-bath sweeps after 5000 thermalization sweeps) with $\beta = 6.0$ on a $20^3 \times 30$ lattice.

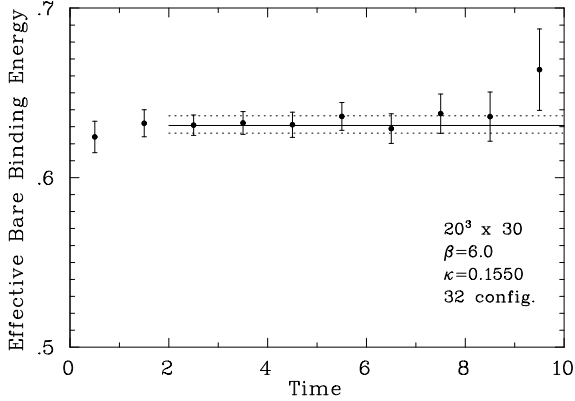

Figure 1. Ground state effective bare binding energy. (Local sink, smeared source).

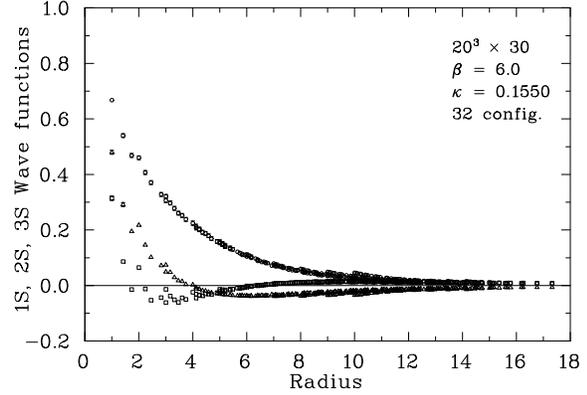

Figure 2. 1S, 2S, and 3S wave functions, normalized to unity at the origin.

Figure 1 displays an effective-$\mathcal{E}^0$ plot for the local-smeared correlation matrix. There is a long plateau beginning at very early times indicating the effectiveness of the procedure in isolating the ground state. Figure 2 displays Bethe-Salpeter amplitudes for the lowest three $S$-wave states, normalized to unity at $\vec{r} = \vec{0}$, and measured using dynamically-improved smeared sources. To obtain values for the decay constant, correlated fits in $t$ and extrapolations in $\kappa$ to $\kappa_c = 0.157$ were made with covariance matrices estimated from single-elimination jackknives (nested when required). Quoted errors are for the bare lattice quantities only, are statistical only, and were obtained by bootstrapping, with 250 samples, the entire fitting procedure. We find provisional values at $\beta = 6.0$ of $\tilde{Z}_L \sqrt{2\kappa} = 0.279^{+6}_{-6}$, $0.244^{+6}_{-5}$, and $0.224^{+6}_{-5}$ at $\kappa = 0.152$, $0.154$, and $0.155$ respectively. With the renormalization factor [5] $\tilde{Z}_A^{\text{stat}} \equiv Z_A^{\text{stat}} \sqrt{2\kappa} \sqrt{4\kappa_c/\kappa - 3}$ with $Z_A^{\text{stat}} = 0.70$, $m_B = 5.28$Gev and a nominal value of $a^{-1} = 2.0$GeV [6], the extrapolation to $\kappa = \kappa_c = 0.157$ is

$$\frac{f_B^{\text{stat}}}{\frac{Z_A^{\text{stat}}}{0.70} \times \left(\frac{a^{-1}}{2.0\text{GeV}}\right)^{3/2}} = 224^{+9}_{-7}(\text{statistical}) \qquad (12)$$

This is in agreement with the results of the Fermilab group who also use nearly perfect sources, but is lower than the results of other groups [7].

In conclusion, the variational basis of MOST generates optimal interpolating fields objectively from the data – no ansatz is needed. These are used to determine wave functions, decay constants and bare binding energies $\mathcal{E}^0$. Effective-$\mathcal{E}^0$ plots plateau early, are flat and have small statistical errors. Results have been obtained for the lowest $S$-wave states; the method generalizes for orbital excitations. The interpolating fields will be optimal choices for use in matrix element calculations of further properties of $B$ mesons.